\documentclass[aps,prl,showpacs,twocolumn,superscriptaddress]{revtex4}

\usepackage{epsfig}
\usepackage{graphicx}
\usepackage{amsmath}
\usepackage{amssymb}
\usepackage{amsfonts}
\usepackage{dcolumn}
\usepackage{color}

\newcommand{\sout}[1]{}

\begin{document}
\title{Bright Solitary Waves in Malignant Gliomas} 

\author{V\'{\i}ctor M. P\'erez-Garc\'{\i}a}
\affiliation{Departamento de
Matem\'aticas, E. T. S. I. Industriales and Instituto de Matem\'atica Aplicada a la Ciencia y la Ingenier\'{\i}a, Universidad de Castilla-La
Mancha, 13071 Ciudad Real, Spain.}

\author{Gabriel F. Calvo}
\affiliation{Departamento de
Matem\'aticas, E. T. S. I. Caminos, Canales y Puertos and Instituto de Matem\'atica Aplicada a la Ciencia y la Ingenier\'{\i}a, Universidad de Castilla-La
Mancha, 13071 Ciudad Real, Spain.}

\author{Juan Belmonte-Beitia}
\affiliation{Departamento de
Matem\'aticas, E. T. S. I. Industriales and Instituto de Matem\'atica Aplicada a la Ciencia y la Ingenier\'{\i}a, Universidad de Castilla-La
Mancha, 13071 Ciudad Real, Spain.}

\author{David Diego}
\affiliation{Departamento de
Matem\'aticas, E. T. S. I. Industriales and Instituto de Matem\'atica Aplicada a la Ciencia y la Ingenier\'{\i}a, Universidad de Castilla-La
Mancha, 13071 Ciudad Real, Spain.}

\author{Luis P\'erez-Romasanta}
\affiliation{Radiation Oncology Service, Hospital General Universitario de Ciudad Real, c/ Tomelloso, s/n, 13005 Ciudad Real, Spain}

\date{\today}

\begin{abstract}
We put forward a nonlinear wave model describing the fundamental physio-pathologic features of an aggressive type of brain tumors: glioblastomas. Our model accounts for the invasion of normal tissue by a proliferating and propagating rim of active glioma cancer cells in the tumor boundary and the subsequent formation of a necrotic core. By resorting to numerical simulations, phase space analysis and exact solutions, we prove that bright solitary tumor waves develop in such systems. 
\end{abstract}


\maketitle

\emph{Introduction.-} Solitons are localized wave-packets able to maintain their shape and speed when propagating in different media and 
under mutual collisions. The existence of such particle-like waves interacting elastically is typical of integrable nonlinear equations~\cite{Scott} but they also arise in a broader set of physical systems described by nonintegrable wave equations, displaying richer spatial interactions with their sustaining media~\cite{calvo} and complex collision scenarios~\cite{super}. In the last years there has been an increased interest in the application of the concepts and tools from nonlinear physics to biology and medicine were nonlinear waves can potentially appear~\cite{Murray}. However, the identification of nonlinear waves in oncology has remained very limited~\cite{Byrne}.
\par
Gliomas comprise a heterogeneous group of neoplasms that initiate in the brain or in the spine. Glioblastoma multiforme (GBM) is the most common and most aggressive type of glioma with poor prognosis and survival ranging from 12 to 15 months after diagnosis~\cite{Wen}. GBMs are composed of a mixture of poorly differentiated neoplastic astrocytes. These tumors may develop (spanning from 1 year to more than 10 years) from lower-grade astrocytomas (World Health Organization [WHO] grade II) or anaplastic astrocytomas (WHO grade III), but more frequently, they manifest {\em de novo}, presenting after a short clinical history, usually less than 3 months, without any evidence of a less malignant precursor lesion~\cite{Meir}. Standard treatments of GBMs include surgery, conformal radiotherapy and drugs such as alkylating agents and antiangiogenic therapies~\cite{Huse}. 
\par

\begin{figure}
\epsfig{figure=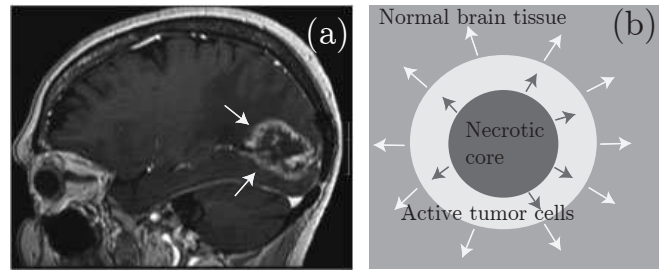, width=1\columnwidth}
\caption{(a) Sagittal magnetic resonance image showing a patient with a GBM in the occipital (posterior) lobe of the brain. The contrast enhanced region (marked by arrows) contains the active tumor cells and the darker inner part of the rim is the necrotic tissue. (b) Schematic representation of the different areas. Arrows indicate the tumor and necrotic core growth directions.
 \label{prima}}
\end{figure}

Various models have been proposed to describe specific aspects of GBMs~\cite{Bondiau,Konukoglu,Rockne,Khain,SwansonB,Eikenberry,Frieboes}, many of them based on a simple reaction-diffusion equation: the Fischer-Kolmogorov (FK) equation~\cite{Murray}. In one-dimensional scenarios the FK equation has solitary wave solutions of kink-type~\cite{Murray,Ablo2}, accounting for the progression of the tumor front edge, but in higher dimensions its analysis must resort to numerical methods. Essential features of high-grade gliomas are neglected by the FK model, such as (i) the formation of a core of necrotic tissue (see Fig.~\ref{prima}), responsible for the intracranial deformation that may lead to death and (ii) the interaction of the tumor with adjacent normal tissue. More elaborated approaches include additional details in (only) some of the intervening mechanisms~\cite{Bondiau,Konukoglu,Rockne,Eikenberry,Frieboes} but lack sufficient biological information on the parameters and  other processes that impair the model's predictive capability. 
\par

There is thus a need for models accounting for the crucial features of GBM dynamics (see Fig.~\ref{prima}) without incorporating excessive details on any of the -often unknown- specific processes. Ideally, such models should be simple enough to allow for some quantitative understanding, e.g. using tools of nonlinear wave theory~\cite{Scott}. In this letter we present a model capturing the key features seen in real GBMs and enabling us to carry out a full theoretical analysis. The model predicts the existence of bright solitary waves, \emph{bright solitons}, for the spreading tumor cell front.
\par

\emph{The model.-} We will consider a simple description of the invasion phenomenon in high-grade gliomas by incorporating the interaction of three relevant (nonnegative) densities. The tumor cell density, denoted by a function $u(x,t)$, the adjacent normal cells $v(x,t)$, and a necrotic core density $w(x,t)$, so that our model reads as
\begin{subequations}
\label{dimensional}
\begin{eqnarray}
\frac{\partial u}{\partial t} & =& D \Delta u + \rho(u_*-u-v-w)u - \alpha u\, , \label{tumor} \\
\frac{\partial v}{\partial t}  & = & -\mathcal{F}(u,v,w)\, , \label{normal} \\
\frac{\partial w}{\partial t}  & = & \mathcal{F}(u,v,w) + \alpha u\, , \label{necrosis}
\end{eqnarray}
\end{subequations}
where $\Delta = \sum_{j=1}^N \partial^2/\partial x_j^2$ and the boundary conditions are $u(\pm \infty) = 0, v(\pm \infty) = v_*, w(\pm \infty) = 0$.
For the tumor to grow it is necessary that $\rho u_{*}>\alpha$. The standard FK equation is recovered by setting $v = w =0 $, however, in order to properly describe the observed clinical phenomenology, we include both the population of the normal tissue and the developing necrotic core, this being a distinctive feature of GBMs (see Fig. \ref{prima}). Here, we will focus in the one-dimensional version of Eqs.~(\ref{dimensional}) but the shown phenomena persist in higher dimensions.
\par

In Eqs.~(\ref{dimensional}), tumor cell spreading is incorporated using a standard Fickian diffusion mechanism. This is the diffusion employed in most of the continuous mathematical models of cell motility. Diffusion phenomena in gliomas should probably be governed by more complicated fractional (anomalous) diffusion~\cite{fractional} or other more elaborate terms~\cite{mechanistic} to account for the high infiltration observed in this type of tumors~\cite{infiltration} and the fact that cells do not behave like purely random walkers and may actually remain immobile for a significant amount of time before compelled to migrate to a more favorable place. The nonlinear term in Eq.~(\ref{tumor}) corresponds to proliferation under a competition for space between the different densities. A tumor cell death term is added to include the fact that tumor cells, although generally lacking the programmed cell death mechanisms~\cite{hallmarks}, may succumb because of their competition with the immune system in normal tissue, hypoxia and acidosis in the high density tumor areas, and deficiency of nutrients and physical support in the necrotic core. In average, the characteristic tumor cell life time is $1/\alpha$. 

In Eq.~(\ref{normal}) we have represented the cell loss due to the interaction with the tumor by means of an arbitrary form $\mathcal{F}(u,v,w)$ depending on all the densities. The details of the interaction may be very complicated. The mechanisms of cell death combine the acidosis generated by the anomalous metabolism of the tumor cells~\cite{Gatenby}, the competition for nutrients with the glioma cells, the modification of the microenvironment~\cite{Annals}, and other effects. In agreement with physiological data, we will assume that the normal brain tissue does not proliferate. Finally, we have assumed in Eq.~(\ref{necrosis}) that the space occupied by the necrotic core is the same compartment occupied by the original cells and grows at the expense of the other two compartments. More elaborate models could include a reduction coefficient to account for the shrinkage of the cells and/or the destruction of their cytoplasm and the release of the cellular content to the necrotic area.
 \par

\begin{figure}[b]
\begin{center}
 \epsfig{figure=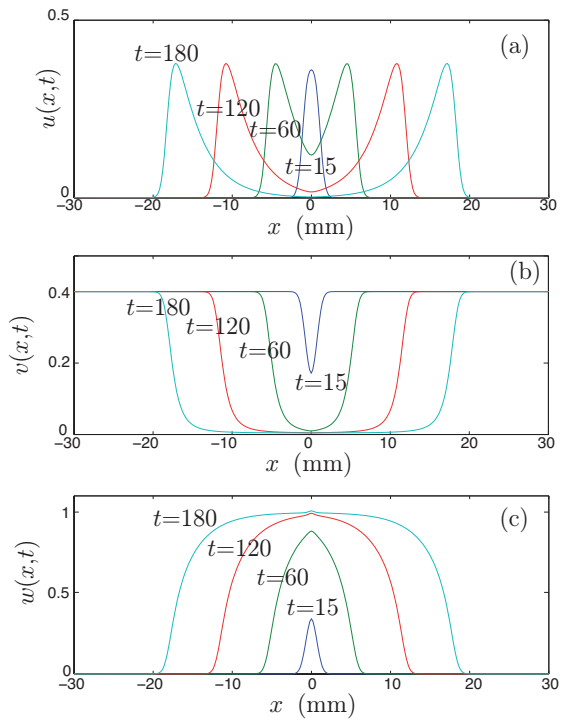, width=0.85\columnwidth}
 \end{center}
\caption{(color online) Simulations of Eqs.~(\ref{dimensional}) in one spatial dimension leading to the formation of a bright soliton. Parameter values are: $D =0.02$ mm$^2$/day, $\rho = 0.5$ day$^{-1}$, $\alpha= 1/30$ day$^{-1}$, and $\gamma = 0.25$ day$^{-1}$. Shown are the profiles of the density of the different densities (a) tumor $u(x,t)$, (b) normal tissue $v(x,t)$, and (c) necrotic core $w(x,t)$, for times $t=15, 60, 120$ and $t=180$ days. All of the densities are measured in units of $u_*$.}
\label{example1}
\end{figure}

\begin{figure}[ht]
\begin{center}
 \epsfig{figure=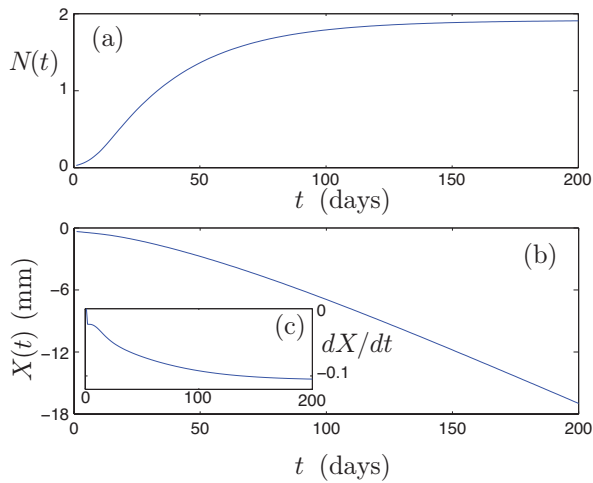, width=0.9\columnwidth}
 \end{center}
\caption{(color online) Simulations of Eqs.~(\ref{dimensional}) in one spatial dimension leading to the formation of a bright soliton. Parameter values are as in Fig.~\ref{example1}.  Evolution of the total cell number (a), center (b), and speed (c) of the soliton traveling to the left, defined as $N(t) =  \int_{-\infty}^0 u(x,t) dx$, $X(t)  = \int_{-\infty}^0 x u(x,t) dx/N(t)$, and $dX/dt$, respectively.}
\label{example2}
\end{figure}

\emph{Numerical simulations.-} To elucidate the typical dynamics of Eqs.~(\ref{dimensional}) we display the results of numerical simulations in Figs.~\ref{example1} and ~\ref{example2}. For this particular case we have chosen a contact interaction term between the tumor and the normal tissue of the form $\mathcal{F}(u,v,w) = \gamma \, u \, v$. The used parameters come from clinically observed values for the diffusion coefficient $D =0.02$ mm$^2$/day, proliferation rate $\rho = 0.5$ day$^{-1}$, death rate $\alpha= 1/30$ day$^{-1}$, and invasion parameter $\gamma = 0.25$ day$^{-1}$. Initial data are taken as $u(x,0) = 0.1 \text{sech}(5x)$, $v(x,0) = v_{*} = 0.4$, $w(x,0) = 0$ (all in units of $u_*$) corresponding to a small dysplasia in a bed of normal cells. 

\begin{figure}
\epsfig{figure=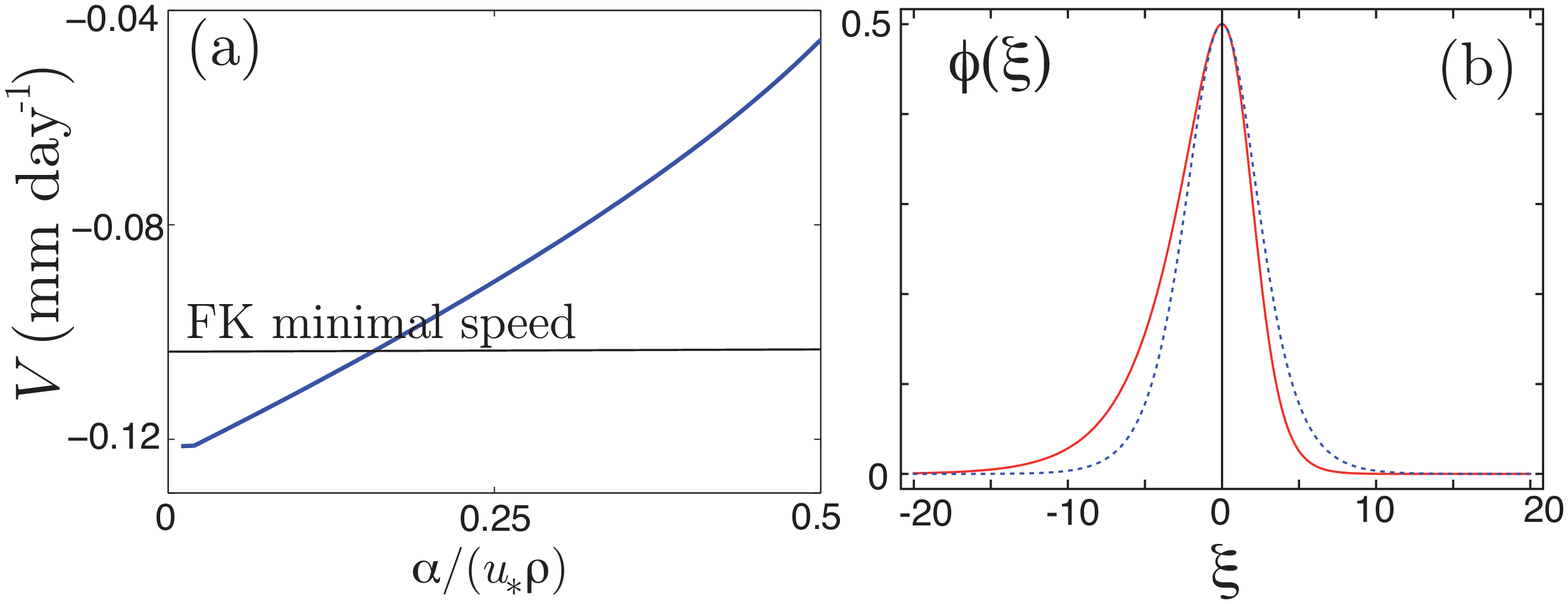, width=1\columnwidth}
\caption{(color online) Bright solitons. (a) Velocity dependence on $\alpha/\rho u_*$. The other parameters and initial data are as in Fig.~\ref{example1}.
(b) Profiles from Eq.~(\ref{autonomous}) (solid curve) and the explicit solution given by Eq.~(\ref{brightsoliton}) (dashed curve) for $\beta = \alpha/\rho u_*= 0.4$, $c=3\sqrt{\rho u_{*} D}$ and $\phi_{0} = 0.5u_{*}$.}
\label{delta}
\end{figure}

Initially ($t=15$ days in Fig.~\ref{example1}), the tumor develops embedded in the environment of normal cells and with space for proliferation. As space saturates because of the accumulation of dead tissue and tumor cells, the cells can no longer survive and start depleting the tumor at its initial location (see e.g. $t=60$ days in Fig.~\ref{example1}). The tumor then generates a bright soliton that propagates in the normal tissue together with a kink in the normal tissue and an anti-kink in the necrotic core (c.f. Fig.~\ref{example1}). After the initial transient the tumor mass stabilizes [see e.g. Fig.~\ref{example2}(a)] following the classical Gompertzian growth curve but now with nontrivial spatial effects incorporated. Finally, as seen in Fig.~\ref{example2}(c), the speed of the soliton stabilizes at a constant value \emph{fundamentally different} from the one from the FK model and dependent on $\alpha$ [Fig.~\ref{delta}(a)]. As with the FK model, faster solitons from initial moving data may be possible, but here we focus our attention on the minimal speed solutions still arising from initial data.

\emph{Theory.-} It is convenient to introduce the new functions $U(\zeta,\tau) = u/u_{*}$ and $S(\zeta,\tau) = 1 - \beta + [v + w]/u_*$, together with the rescaled variables $\zeta = x\sqrt{\rho u_*/D}$ and $\tau = \rho u_{*}t$. Then, Eqs. (\ref{dimensional}) can be cast as
\begin{subequations}
\label{adimensional}
\begin{eqnarray}
U_{\tau} & =& U_{\zeta\zeta} - (U+S)U\, , \label{tumor-adim} \\
S_{\tau} & = & \beta U\, , \label{normal-adim}
\end{eqnarray}
\end{subequations}
with $\beta = \alpha/\rho u_{*} < 1$ and subscripts will henceforth denote partial differentiation. Let us first notice that, surprisingly, the precise form of the tumor-host interaction term $\mathcal{F}$ is not relevant for the $U$-$S$ dynamics. Instead, the global function $S$ incorporating both the necrotic tissue and the normal cells accounts for the effect of the peritumoral environment. Secondly, because of Eq.~(\ref{normal-adim}), the non-tumoral density at any given point always increases which implies that after the tumor wave crosses a region, the necrotic core contains a higher density than the normal stroma. Thirdly, since the physically feasible solutions must be positive, we combine Eqs.~(\ref{adimensional}) into a single (and more general) equation for the tumor density
\begin{eqnarray}
\left[ \frac{1}{U}\left( U_{\tau} - U_{\zeta\zeta} + U^{2}\right)\right]_{\tau} = -\beta U\, .
\label{single}
\end{eqnarray}

\emph{Short time limit.-}  When $0< \tau\ll 1/\beta$, the right-hand-side term in Eq.~(\ref{single}) can be neglected (no significant tumor cell death has occurred yet). The resulting equation $U_{\tau} - U_{\zeta\zeta} + U^{2} = 0$ possesses self-similar solutions of the form $U(\zeta,\tau)=\varphi\left( \zeta\tau^{-1/2}\right)/\tau$, where $\varphi$ satisfies $2\left( \varphi_{\eta\eta} + \varphi -\varphi^{2}\right) + \eta\varphi_{\eta}=0$ and $\eta = \zeta\tau^{-1/2}$. For $U$ to remain constant in the limit $\tau\to0^{+}$, it must be the case that $\varphi\to0$ as $\eta\to\infty$. Hence, we can disregard the nonlinear term, and consider $2\left(\varphi_{\eta\eta} + \varphi\right) + \eta\varphi_{\eta}=0$, whose physically meaningful solution is $\varphi=\varphi_{0}\vert\eta\vert\exp(-\eta^{2}/4)$. This profile is consistent with the initial ones observed in the numerical simulations [see Fig.~\ref{example1}(a)].

\emph{Solitary waves.-} Our numerical simulations show that two counter-propagating tumor wave fronts develop for $t\gg 1/\alpha$. In this long time limit, both fronts acquire a characteristic traveling solitary shape (they no longer interact). We will look for such solutions of Eq.~(\ref{single}) in the form $U(\zeta -c\tau) = \phi(\eta)$, where $c$ is their velocity (positive or negative for right- or left-moving fronts, respectively). These solutions will be required to satisfy $\phi < 1-\beta$, together with $\phi\to 0$ and $\phi_{\xi}\to 0$ as $\vert\eta\vert\to\infty$. Upon substitution in Eq.~(\ref{single}) and defining $\xi = \eta/c$, we arrive at a third order nonlinear autonomous differential equation
\begin{eqnarray}
\beta\phi^{3}  -\phi_{\xi}^{2} - \phi^{2}\phi_{\xi} + \phi\phi_{\xi\xi} - \frac{1}{c^{2}}(\phi_{\xi}\phi_{\xi\xi} -\phi\phi_{\xi\xi\xi})= 0\, .
\label{autonomous}
\end{eqnarray}
The existence of traveling waves of Eq.~(\ref{autonomous}) as a function of $\beta$ and $c$ is a mathematical bifurcation problem beyond the scope of this paper. Our numerical calculations reveal that: (i) the minimum (absolute value) speed $c_\textrm{min}$ for which positive solutions $\phi <1-\beta$ exist decreases as $\beta\to 0$ and can attain values $c_\textrm{min}<2\sqrt{\rho u_{*} D}$. This is in contrast with the FK equation for which the minimum speed must always fulfill $c_\textrm{min}\geq 2\sqrt{\rho u_{*} D}$~\cite{Murray}. Therefore, our model [see Fig.~\ref{delta}(a)] can account for slower growing gliomas than those predicted by the standard FK equation~\cite{SwansonB}. (ii) For $\vert c\vert>5\sqrt{\rho u_{*} D}$, the last two terms in Eq.~(\ref{autonomous}) can be neglected for any $\phi <1-\beta$ to yield the simpler equation
\begin{eqnarray}
\beta\phi^{3}   -\phi_{\xi}^{2} - \phi^{2}\phi_{\xi} + \phi\phi_{\xi\xi} = 0\, .
\label{autonomous2}
\end{eqnarray}
\emph{Phase portrait analysis.-} Bright soliton solutions to Eq.~(\ref{autonomous2}) correspond to homoclinic paths in the phase plane.
Defining $\psi(\phi)=\phi_\xi$, Eq.~(\ref{autonomous2}) transforms into the orbit equation $\psi_\phi=\phi+\psi/\phi-\beta \phi^2/\psi$,
whose solution, with $\chi = \psi/\phi$, is $\chi+\beta\,{\rm log}\left|\beta -\chi\right| =\mathcal{C} +\phi$. The critical points along the orbits obey $\phi=\chi^2/(\beta-\chi)$ giving rise to two distinct regions for $\phi>0$. The region $\chi>\beta$ contains no localized solutions since the orbit $\psi(\phi)$ is a monotonic increasing function. The region $\chi<\beta$ does contain localized solutions if the integration constant $\mathcal{C}$ is bounded from above. This can be shown by considering
\begin{equation}
f(\chi) = \chi+\beta\,{\rm log}\left(\beta-\chi\right) -\frac{\chi^2}{\beta-\chi}-\mathcal{C}\,,
\label{Auxiliaryfunction}
\end{equation}
which presents a global maximum at $\chi=0$. In addition, it holds that $\lim_{\chi\to-\infty} f(\chi) = \lim_{\chi\to\beta^{-}} f(\chi) = -\infty$. Thus, the existence of two different roots (with opposite signs) is guaranteed whenever $\mathcal{C}<\beta\,{\rm log}\beta$. This proves the existence of fast bright soliton solutions. A complete theoretical analysis will be reported elsewhere.
\par 
\emph{Explicit soliton solutions.-} As a final comment, let us mention that the approximate explicit form of bright soliton solutions from Eq.~(\ref{autonomous2}) can be obtained by neglecting the third term in  Eq.~(\ref{autonomous2}). This term introduces a small asymmetry in the profile. The found expression is
\begin{eqnarray}
\phi(\xi) = \phi_{0}\,\textrm{sech}^{2}\!\left[ \sqrt{\frac{\phi_{0}\beta}{2}}(\xi - \xi_{0})\right],
\label{brightsoliton}
\end{eqnarray}
with constants $\phi_{0}$ and $\xi_{0}$. Fig.~\ref{delta}(b) compares the exact numerical profile from Eq.~(\ref{autonomous}) and the one predicted by Eq.~(\ref{brightsoliton}). The non-tumoral density $S$ can  be obtained by integrating Eq.~(\ref{normal-adim}) and exhibits a kink-like shape. Within the necrotic compartment, this is the typical contrast-enhanced region seen in the magnetic resonance images. The tumor cell number $N(t) = \int_{-\infty}^{0} u(x,t) dx$ for the left-moving soliton follows from Eq.~(\ref{brightsoliton})
\begin{eqnarray}
N(t) = \sqrt{\frac{2u_{*}^{2}\phi_{0}c^{2}D}{\alpha}}\left[ 1 + \tanh\left( \sqrt{\frac{\phi_{0}\rho u_{*}\alpha}{2}}\,t\right)\right] ,
\label{totalnumber}
\end{eqnarray}
which, for long times, gives a saturation similar to the one depicted in Fig. \ref{example2}(a). 

\par

\emph{Conclusions.-} We have developed a simple model of glioblastoma progression incorporating the normal tissue, tumor cells and the necrotic core. Our theoretical study displays many of the signatures of aggressive malignant gliomas with bright solitons acting as attractors of the tumor-host dynamics that can be compared with the observed phenomenology. The two and three dimensional versions of our starting Eqs.~(\ref{dimensional}) can easily accommodate the effect of ionizing radiation both on the tumor and on the normal tissue~\cite{radiobiology}. This could provide a useful tool, if combined with intensity-modulated radiation therapy, to simulate the outcome of different dose painting scenarios. We hope that our model will stimulate the use of techniques from nonlinear physics and nonlinear wave theory to further understand key aspects of tumor growth.

\acknowledgments

\emph{Acknowledgements.-} This work has been supported by grants MTM2009-13832 (Ministerio de Ciencia e Innovaci\'on, Spain) 
and PEII11-0178-4092 (Junta de Comunidades de Castilla-La Mancha, Spain). We thank A. Mart\'{\i}nez, B. Mendoza and M. Ortigosa for discussions.


\begin{thebibliography}{99}

\bibitem{Scott}  A. C. Scott, \emph{The Nonlinear Universe: Chaos, Emergence, Life}, Springer (2007).

\bibitem{calvo} G.F. Calvo, B. Sturman, F. Agullo-Lopez and M. Carrascosa, Phys. Rev. Lett. \textbf{89}, 033902 (2002).

\bibitem{super} D. Novoa, B. A. Malomed, H. Michinel and V. M. P\'erez-Garc\'{\i}a, Phys. Rev. Lett. \textbf{101}, 144101 (2008).

\bibitem{Murray} J. Murray, \emph{Mathematical Biology}, Third Edition, Springer (2007).

\bibitem{Byrne} H.M. Byrne, Nature Rev. Cancer \textbf{10}, 221 (2010).

\bibitem{Wen} P.Y. Wen and S. Kesari, New Eng. J. Med. \textbf{359}, 492 (2008).

\bibitem{Meir} E.G.V. Meir, \emph{et al.},
C.G. Hadjipanayis, A.D. Norden, H.-K. Shu, P.Y. Wen and J.J. Olson, 
CA Cancer J. Clin. \textbf{60}, 166 (2010).

\bibitem{Huse} J.T. Huse and E.C. Holland, Nature Rev. Cancer \textbf{10}, 319 (2010).

\bibitem{Khain} E. Khain and L.M. Sander, Phys. Rev. Lett. \textbf{96}, 188103 (2006).

\bibitem{SwansonB} K.R. Swanson, R. C. Rostomily and E. C. Alvord Jr, British J. Cancer \textbf{98}, 113 (2008).

\bibitem{Frieboes} H. B. Frieboes, \emph{et al.}, 
Neuroimage \textbf{37}, S59 (2007).

\bibitem{Bondiau} P.-Y. Bondiau {\em et al}, 
Phys. Med. Biol. \textbf{53}, 879 (2008).

\bibitem{Eikenberry} S. E. Eikenberry {\em et al}, 
Cell Prolif. \textbf{42}, 511 (2009).

\bibitem{Konukoglu} E. Konukoglu, O. Clatz, P.-Y. Bondiau, H. Delingette and N. Ayache, Med. Image Anal. \textbf{14}, 111 (2010).

\bibitem{Rockne} R. Rockne {\em et al}, 
 Phys. Med. Biol. \textbf{55}, 3271 (2010).


\bibitem{Ablo2} M. J. Ablowitz and A. Zeppetella, Bull. Math. Biol. \textbf{41}, 835 (1979); Y. N. Kyrychko and K. B. Blyuss, Phys. Lett. A \textbf{373}, 668 (2009).

\bibitem{fractional} D. \ del-Castillo-Negrete, B. A. Carreras and V. E. Lynch, Phys. Rev. Lett. \textbf{91}, 018302 (2003); S. Fedotov and A. Iomin, Phys. Rev. Lett. \textbf{98}, 118101 (2007). 

\bibitem{mechanistic} C. Deroulers, M. Aubert, M. Badoual and B. Grammaticos, Phys. Rev. E \textbf{79}, 031917 (2009); M. J. Simpson, A. Merrifield, K. A. Landman and B. D. Hughes, Phys. Rev. E \textbf{76}, 021918 (2007).

\bibitem{infiltration} M. Onishi, T. Ichikawa, K. Kurozumi and I. Date, Brain Tumor Pathol. \textbf{28}, 13 (2011).

\bibitem{hallmarks} D. Hanahan and R.A. Weinberg, Cell \textbf{144}, 646 (2011).

\bibitem{Gatenby} R. Gatenby and E. Gawlinski, Cancer Res. \textbf{56}, 5745 (1996).

\bibitem{Annals} N. E. Savaskan and I. Y. Eyupoglu, Ann. Anatom. \textbf{192}, 309 (2010).

\bibitem{radiobiology} M. Joiner and A. van der Kogel (eds.), \emph{Basic Clinical Radiobiology}, Fourth Edition, Hodder Arnold (2009).

\end{thebibliography}
\end{document}